\documentclass[AMA,STIX1COL]{WileyNJD-v2}

\articletype{}%

\received{}
\revised{}
\accepted{}
        
\begin{document}

\title{A new rank-order clustering algorithm for prolonging the lifetime of  wireless sensor networks}

\author[1]{SeyedAkbar Mostafavi*}
\author[2]{Vesal Hakami}

\authormark{AUTHOR ONE \textsc{et al}}

\address[1]{\orgdiv{Department of Computer Engineering}, \orgname{Yazd University}, \orgaddress{\state{Yazd}, \country{Iran}}}
\address[2]{\orgdiv{Department of Computer Engineering}, \orgname{Iran University of Science and Technology}, \orgaddress{\state{Tehran}, \country{Iran}}}

\corres{* \email{a.mostafavi@yazd.ac.ir}}

%\presentaddress{This is sample for present address text this is %sample for present address text}

\abstract[Summary]{Energy efficient resource management is critical for prolonging the lifetime of wireless sensor networks (WSN). Clustering of sensor nodes with the aim of distributing the traffic loads in the network is a proven approach for balanced energy consumption in WSN. The main body of literature in this topic can be classified as hierarchical and distance-based clustering techniques in which multi-hop, multi-level forwarding and distance-based criteria, respectively, are utilized for categorization of sensor nodes. In this study, we propose the Approximate Rank-Order Wireless Sensor Networks (ARO-WSN) clustering algorithm as a combined hierarchical and distance-based clustering approach. ARO-WSN algorithm which has been extensively used in the field of image processing, runs in the order of O(n) for a large data set, therefore it can be applied on WSN. The results shows that ARO-WSN outperforms the classical LEACH, LEACH-C and K-means clustering algorithms in the terms of energy consumption and network lifetime. Energy consumption in the proposed approach is about \%10 lower than the LEACH, LEACH-C and K-means algorithms and about \%3 lower than the LEACH algorithm with FUZZY descriptors. The lifetime of the network with the first node death criterion improved relative to LEACH, LEACH-C and LEACH with fuzzy descriptors by \%60, \%85 and \%22, respectively and with last node death criterion improved relative to K-means, LEACH, LEACH-C and LEACH with fuzzy descriptors by \%42, \%67, \%64 and \%24, respectively.}


\maketitle

\section{INTRODUCTION}
Wireless sensor networks have used for a wide range of applications such as environmental monitoring, military monitoring, health care and crisis management \cite{1}. In these environments without supervision, the sensors can be easily replaced or recharged \cite{2}; therefore, energy conservation in the networks is one of the most important research challenges. One method commonly used to prolong the network lifetime is through collecting data at the cluster heads \cite{3}. In cluster-based approaches, sensors do not need to communicate directly with the base station. Instead, cluster heads are responsible to organize cluster members and send the data collected within the cluster to the base station. This process leads to a significant reduction in the amount of data transmitted, resulting in lower energy consumption and longer network lifetime \cite{3}.

LEACH \cite{4} is the first hierarchical routing protocol that is more efficient over traditional routing protocols. In LEACH, the cluster head is selected with a probabilistic method and attempts to distribute the network load at each sensor node \cite{5}. This protocol does not guarantee the number and position of cluster heads. LEACH-C \cite{6} is another routing protocol that follows a centralized approach to elect a cluster head using base station and location information of each node. This will produce better number of clusters and distributes the CHs evenly among the clusters. This method increases the network overhead, since all sensor nodes must send their location information to base station in each round \cite{5}. The K-means clustering algorithm in addition to extensive applications in various domains, such as data mining and image processing, has been used in \cite{7} to cluster and extend the lifetime of wireless sensor networks. The K-means algorithm is based largely on Euclidean distances and selection of clusters dependent on the remaining energy of the nodes \cite{7}. In this algorithm, the parameter K is chosen by the user, which in practice cannot be assumed that the actual number of clusters is predetermined. 

Hierarchical clustering (such as LEACH and LEACH-C) and distance-based clustering (such as K-means) are methods employed to optimize energy consumption and increase the lifetime of wireless sensor networks. The combination of these two methods of clustering is the main motive for this study. Approximate Rank-Order is a hierarchical clustering based on the distance criterion used in image processing to clustering millions of images \cite{8}. From here on, we use the ARO shortcut name to refer to the Approximate Rank-Order clustering. In this paper, a clustering algorithm in wireless sensor networks is provided using ARO clustering, in which the clustering step is performed before selecting the cluster head. The idea that the clustering stage should first be done is based on the work done in \cite{9}. This allows energy consumption in the network to be reduced by eliminating overhead advertisements (which is in all four other algorithms) per round and thus increasing the lifetime of the network. Comparing the time complexity of the proposed approach with other algorithms, the time complexity of this approach in the setup phase and cluster formation is equal to the time complexity of the Approximate Rank-Order algorithm in \cite{8} and equal to O(n). The time complexity of the LEACH algorithm according to \cite{10} is equal to O (1), and LEACH-C and LEACH with fuzzy descriptors algorithms are also O (1). The time complexity of the K-means algorithm is also equal to O (ncdi), in which n is the number of nodes, c is the number of clusters, d is the number of dimensions, and i is the number of repetitions \cite{11}. It is important to note that in the proposed approach, the clustering stage is performed only once, and the complexity of the time O (n) does not have a significant effect on the overall execution time. Because the total execution time for other algorithms is proportional to the number of rounds, in these experiments the number of rounds is greater than the number of nodes.

The remainder of the article is as follows: Section 2 provides a summary of related work in this field. Section 3 introduces the radio model used in this paper. Section 4 describes the proposed approach. Simulation results and comparison of proposed approach with LEACH, LEACH-C, K-means and LEACH with fuzzy descriptors algorithms are presented in Section 5. 

\section{Related Work}

Yarinezhad, Hashemi WSN \cite{yarinezhad2019routing}
LBCH \cite{al2018lbch}
CRDP \cite{wang2018crpd}
Cluster head selection \cite{sarkar2019cluster}
BPA-CRP \cite{darabkh2019bpa}
Fuzzy \cite{hamzah2019energy}
secure routing \cite{maitra2019cluster}
unequal cluster \cite{yang2018unequal}
cluster routing \cite{tuna2018clustering}
mixed integer \cite{li2018clustering}
pareto \cite{elhabyan2018pareto}
qos aware \cite{yahiaoui2018energy}
lifetime \cite{arjunan2018lifetime}
integrated clustering \cite{gupta2018integrated}
CREEP \cite{dutt2018cluster}
neuro fuzzy \cite{thangaramya2019energy}
Bayesian \cite{tyagi2015bayesian}
LA-MHR \cite{tanwar2018mhr}
EEMHR \cite{zhang2019energy}
systematic \cite{tanwar2015systematic}
lifte extended \cite{tyagi2015lifetime}
cognitive \cite{tyagi2015cognitive}
learning automata \cite{tyagi2015learning}
EHE-LEACH \cite{tyagi2013ehe}

Over the past decade, research activities in the field of clustering wireless sensor networks have grown significantly. A large number of clustering algorithms are proposed in wireless sensor networks; Such as LEACH \cite{4}, LEACH-C \cite{6}, K-means \cite{7}, LEACH with fuzzy descriptors \cite{5}, and SCRC-WS \cite{9}. In this section, we provide a brief review on the most popular WSN clustering protocols to get deeper understanding from their functions. This analysis helps us to analyze the pros and cons of state-of-the-art clustering protocols for WSN.

\subsection{LEACH Protocol}
Low Power Adaptive Hierarchy Clustering (LEACH) \cite{4} is a self-organizing adaptive protocol that uses a distributed clustering algorithm \cite{2}. The purpose of LEACH is to randomly select sensor nodes as cluster heads so that energy dissipation in relation to the base station spreads across all sensor nodes in the network. Each clustering cycle consists of the setup phase (cluster formation) and the steady phase (data transfer). During the setup phase, a sensor node selects a random number between 0 and 1. If this random number is less than the threshold T(n), the sensor node is selected as the cluster head. T(n) is calculated as follows \cite{12}:

\begin{align}\label{vequ1}
T(n)=\left\{\begin{array}{lc}\frac P{1-P\;\ast\;\lbrack r\;mod\;({\displaystyle\frac1P})\rbrack}&if\;n\in G\\0&otherwise\end{array}\right.
\end{align}

In equation (1), P is the desired percentage of nodes to become a cluster head, r is the number of rounds for the election, and G is the set of nodes that have not been selected yet as a cluster head in the last round of 1/P. The cluster head nodes notify other nodes after they have been selected. When the sensor nodes receive an advertisement message, they determine the cluster to which they want to belong based on the signal strength of the received message. Then, the cluster heads allocate the time on which the sensor nodes can send data to the cluster heads based on a TDMA approach. During the steady-state phase, the sensor nodes can send data to cluster heads. The cluster heads collect data from the nodes in their cluster before sending these data to the base station. After a certain period of time spent on the steady-state phase, the network re-enters the setup phase and enters another round of selection of cluster heads \cite{12}.

\subsection{LEACH-C Protocol}
While using LEACH has the benefits of a random, adaptive, and self-organized clustering algorithm, this protocol does not guarantee the number and placement of cluster head nodes. Therefore, using a central control algorithm to form clusters will help to produce better clusters by dispersing cluster head nodes throughout the network. This is the core idea behind the LEACH-C protocol which, unlike LEACH, uses a centralized clustering algorithm \cite{6}. The purpose of LEACH-C protocol is to combine energy efficient clustering and routing for application-specific data aggregation to increase network lifetime and reduce latency in data access \cite{13}. In the setup phase of LEACH-C, each node sends information about its current position (possibly determined by the GPS receiver) and current energy level to the base station. In addition to forming good clusters (for example, the proper physical location and the number of nodes that are roughly the same), base station must ensure that the energy load is uniformly distributed over all nodes. To this end, the base station calculates the average energy level of the network nodes so as nodes whose energy levels are below the average cannot be selected as cluster heads for the current round. This algorithm attempts to minimize the amount of energy for nodes that are not cluster heads to transmit their data to the cluster head, by minimizing the total sum of squared distances between all non-cluster head nodes and closest to the cluster head. After determining the cluster heads of the current round, the base station broadcasts a message containing the cluster head ID to each node. If a node’s cluster head ID matches its own ID, the node is a cluster head; otherwise, it is a normal node and determines its TDMA slot for data transmission and goes to sleep until it is time to transmit data \cite{6}. This method increases the network overhead because all sensor nodes must send their location information to each base station at a time in every setup phase \cite{5}.

\subsection{K-means Algorithm in WSN}
Due to the widespread use of the K-means algorithm in various domains, such as data mining and image processing, the authors in \cite{7} have used this method to cluster the sensor nodes and extend the lifetime of wireless sensor networks. The K-means algorithm is mainly based on the Euclidean distance of nodes and selection of cluster heads based on the residual energy of the nodes. In this method, the sink node collects information about the identifier, position, and residual energy of all nodes and stores this information in a list in the central node. After receiving this information from all the nodes, it begins to cluster the sensors using the K-means algorithm. The steps in the K-means clustering algorithm in wireless sensor networks are as follows \cite{7}: 

  \begin{enumerate}
  \item \relax To form the cluster ’k’ of the sensor nodes, the ’k’ number of centroids is initially selected at random locations;
  \item \relax Euclidian distance is calculated from each node to all centroids and allocated to the centroid nearest. Thus the ’k’ initial clusters are formed;
  \item \relax The positions of the centroids in each cluster are recalculated and the position change from the previous one is examined;
  \item \relax If there is a change in the position of each centroid, all steps are repeated from step 2, otherwise, the clusters will be finalized and the clustering process will end.
  \end{enumerate}

\subsection{Introducing the FUZZY-LEACH algorithm}
LEACH Clustering is the most famous hierarchical protocol in which the cluster head is selected based on the threshold value, and only the cluster heads can send the information to the base station. But in this approach, a privileged cluster is selected among the cluster heads, which can send information to the mobile base station by choosing appropriate fuzzy descriptors such as residual energy, mobility, and centrality of clusters. Fuzzy inference engine (Mamdani's rule) is used to select the chance to be superior cluster head \cite{5}.
 Fuzzy Logic is used to model human experience and human decision-making behavior. In addition, it can handle the uncertainty of real-time applications more accurately than probabilistic models. FL is used in this method to control the uncertainty for choosing the superior cluster head. The main advantage of using FL is to overcome the overhead of collecting and calculating the energy and location information of each node. Most FL-based clustering algorithms are considered constant, sink  node / base station \cite{5}.

\section{Energy Model}

\begin{figure}[!ht]
\centering
\includegraphics[width=8cm]{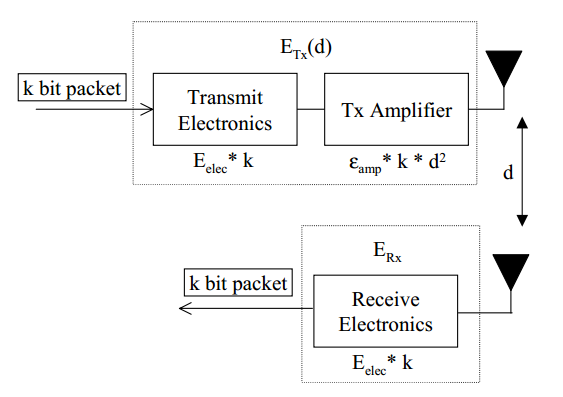}
\caption{The radio model used to calculate energy consumption \cite{4}}\label{vfig1}
\end{figure}

In this section, we will introduce the radio model used in this paper to compare clustering algorithms. We consider the radio model used in \cite{4} for the LEACH protocol. The energy dissipation model is shown in \ref{vfig1}. In Figure \ref{vfig1}, $E_{Tx}(d) $ is the energy spent to transmit k bits over d, and $E_{Rx}$ represents the energy spent to process k bit message. The parameter $E_{elec} $ shows the energy consumed in each bit to run both transmitter and receiver circuits. This parameter depends on many factors such as digital coding, modulation, filtering, and signal propagation \cite{9}.  The energy consumed by the radio transmitter is defined by the following equation \cite{6}:

\begin{align}\label{vequ2}
E_{Tx}(k,\;d)=\left\{\begin{array}{lc}k.E_{elec}+k.E_{fs}.d^2&if\;d<d_o\\k.E_{elec}+k.E_{mp}.d^4&if\;d\geq d_o\end{array}\right.
\end{align}

Where $E_{fs}$ and $E_{mp}$ represent the amplifier energy respectively in the model of free space channel (energy loss $d^2$) and in the model of multipath fading channel (energy loss $d^4$) multipath. They are dependent on d, where d is the distance between the sender and the receiver. If the distance d is less than the threshold $d_{o}$, then the free space model is used; otherwise, the multimode model is used \cite{9}. The threshold value $d_{o}$ is presented by Heinzelman et al. in \cite{6}. $d_{o}$ is defined as follows:

\begin{align}\label{vequ3}
d_o=\sqrt{\frac{E_{fs}}{E_{mp}}}
\end{align}

\section{The Proposed Approach}

In this paper, we adopt the Approximate Rank-Order (ARO) clustering algorithm, which is recently proposed in \cite{8}, as an energy-efficient scheme to cluster nodes in wireless sensor networks. In this section, we first introduce Rank-Order and Approximate Rank-Order algorithms and then, we present our proposed clustering approach. 

\subsection{Introduction of Rank-Order Clustering Algorithm}
The Rank-Order clustering algorithm was proposed by Zhu et al. \cite{14}, which is an agglomerative hierarchical clustering algorithm based on nearest neighbor distance measure. The algorithm works as follows: initially, all the samples are considered as separate clusters, then the distance vector consisting the distance between all the pairs of clusters is computed and the pairs whose distance is below the threshold are merged together. This way onwards, new vector of cluster-to-cluster distance is repeatedly recomputed and merges are performed based on the new distance vector. This main point in Rank-Order clustering is the definition of cluster-to-cluster distance metric. In this algorithm, the distance between two clusters is defined as the minimum distance between any two samples in the clusters. The first measurement of the distances used in Rank-Order clustering is obtained by the following equation \cite{8}:

\begin{align}\label{vequ4}
d(a,\;b)=\sum_{i=1}^{O_a(b)}O_b(f_a(i))
\end{align}

Where $f_a (i)$ is the $i$-th face in the neighbor list of a, and $O_b (f_a (i))$ is the rank of face $f_a (i)$ in the neighbor list of b, which the nearest neighbor lists are generated according to some underlying distance measure (e.g., the Euclidean distance). This asymmetric distance function is used to define a symmetric distance between two faces a and b. as \cite{8}:

\begin{align}\label{vequ5}
D(a,\;b)=\frac{d(a,\;b)+d(b,\;a)}{min(O_a(b),\;O_b(a))}
\end{align}

A cluster-level normalized distance measure is effectively used to restrict mergers to local neighborhoods. Specifically, the minimum distance between the two points in a pair of clusters is calculated and divided by the mean distance \cite{7873333}:

\begin{align}\label{vequ6}
D^N(C_i,\;C_j)=\frac1{\phi(C_i,\;C_j)}\times d(C_i,\;C_j)
\end{align}

Where

\begin{align}\label{vequ7}
\varphi(C_i,\;C_j)=\frac1{\mid C_i\mid +\mid C_j\mid }\times\sum_{a\in C_i\cup C_j}\frac1K\sum_{k=1}^Kd(a,\;f_a(k))
\end{align}

Where $D^N (C_i,C_j )$, is the minimum distance between any two points in the cluster i and j divided by the mean distance of each point in $C_i$ or $C_j$ to their K nearest neighbors. The threshold 1 is always used for this function, and it is tested whether or not the minimum distance between any two points in the clusters is below the average distance of all points in both clusters to their K nearest neighbors \cite{8}.

The symmetric rank-order distance function returns low values if the two samples are close to each other and have neighbors in common. After calculating the distances, clustering phase begins where each sample is considered as a distinct cluster, then the symmetric distances between each pair of clusters are calculated, and finally, clusters with the distance below the specified threshold are merged together. Afterwards, the nearest neighbor lists are updated for each newly formed cluster, and the distances between the remaining clusters are recalculated; this approach continues until no further clusters can be merged. In Rank-Order algorithm, instead of determining the optimal number of clusters C, a distance threshold is used to measure the rank-order distance along with a neighborhood size for the cluster-level normalized distance; these parameters set the specified number of clusters for a particular dataset, and their effective values are determined empirically \cite{8}.

In terms of runtime, calculation the full nearest neighbor lists for each sample affords $O(n^2)$ cost. Additionally, the actual clustering step used here is iterative, with cost per iteration proportional to the current number of clusters squared, so both the nearest neighbor computation and the clustering step itself are costly with increasing dataset size \cite{8}.

\subsection{Review of ARO algorithm}
The Rank-Order clustering method has an obvious scalability problem that requires calculating the nearest neighbor lists for each sample in the data set, which if calculated directly, has an $O(n^2)$ cost. Although there are various approximate methods for calculating the nearest neighbors, they are usually only able to calculate a short list of the top k nearest neighbors rather than a comprehensive ranking the dataset \cite{8}.
Using approximate methods for faster calculation of the nearest neighbor requires some modification in the original Rank-Order clustering algorithm. In particular, instead of taking into account all the neighbors in the summation equation (\ref{vequ4}), we can sum up to at most the top k neighbors. In addition, instead of using the cluster-level normalized distance from the original algorithm, it is possible to approximate the distance between the pairs of samples for which both are within the other sample’s top-200 nearest neighbors. It should be noted that if only one short list of the top-k neighbors is considered, the presence or absence of a particular instance in this list may be more important than the sample’s numerical rank. Thus, a distance measure based on the sum of the presence/absence of the shared nearest neighbors, instead of the ranks, can be used which results in the following distance function \cite{8}:

\begin{align}\label{vequ8}
d_m(a,\;b)=\sum_{i=1}^{min(O_a(b),\;k)}I_b(O_b(f_a(i)),\;k)
\end{align}

Where $I_b (x,k)$ is an indicator function with the value of 0 if face x is in face b’s top k nearest neighbors, otherwise it is equal to 1. In practice, this modification leads to better clustering accuracy compared to the sum of ranks in a direct way, as in the original formula. Effectively, this distance function indicates that the presence or absence of shared neighbors towards the top of the nearest neighbor list (within the top-200 ranks) is important, while the numerical values of ranks themselves are not \cite{8}.

The normalization method used in the original algorithm is still effective and contributes to more accurate clustering results even with this modification to the original algorithm. The modified distance measure is defined as follows \cite{8}:

\begin{align}\label{vequ9}
D_m(a,\;b)=\frac{d_m(a,\;b)+d_m(b,\;a)}{min(O_a(b),\;O_b(a))}
\end{align}

In addition, in order to improve the runtime of the clustering process, as mentioned earlier, 1) only the distances between samples which appear in each other’s nearest neighbor lists are calculated, and 2) instead of multiple merge iterations as in the original algorithm, only one round of merges of individual faces into clusters are done. This means that only one clustering repetition is performed compared to the original algorithm that has runtime $C^2$ in each cluster iteration, and additionally only check for merges on a subset of all possible pairs (because only 200 nearest neighbors for each sample Is considered). This results in the runtime of $O(n)$ from the final clustering step (assuming that the nearest neighbors have been pre-calculated) \cite{8}.

Finally, the clustering approach in ARO algorithm is as follows \cite{8}:

  \begin{enumerate}
  \item \relax Extract dataset;
  \item \relax Calculate a set of top-k nearest neighbors for each face in the dataset;
  \item \relax Calculate pairwise distances between each face, and those faces in its top-k nearest neighbor list for which the face is also on the neighbor’s nearest neighbor list, following equation (\ref{vequ9});
  \item \relax Merge all pairs of faces with distances below a threshold.
  \end{enumerate}

Choosing the threshold to determine the number of clusters, C in a dataset, is one of the most difficult issues of data clustering. In practical applications, it cannot be assumed that the actual number of clusters is predetermined, so this algorithm is evaluated in several effective values of C and the best result is reported \cite{8}.

\subsection{Proposed ARO-WSN approach}
In this paper, we intend to use the ARO algorithm in order to cluster nodes in a wireless sensor environment. In addition to applying ARO algorithm in the clustering step, we will also take advantage of the idea employed in \cite{9} where the clustering step is performed before selecting cluster heads. Our proposed approach considers the residual energy of nodes when selecting the cluster heads. In the following, three steps of the proposed algorithm including clustering, cluster head selection and steady-state will be described.

\subsubsection{Clustering  step }
At this point, network sensors will cluster based on the ARO clustering algorithm. The algorithm uses a distance measurement based on the sum of the presence/absence of shared nearest neighbors, instead of the ranks that are formulated in equation (8). The final clustering method we use is the same as the ARO clustering steps, with the difference that in the third step, instead of using the cluster-level normalized symmetric distance, we use the same asymmetric equation (8) to calculate the distance between the sensor nodes. Therefore, the clustering step of the proposed approach is as follows:

\begin{enumerate}
  \item \relax Collect the location of the sensor nodes by the base station. Each sensor node knows its location (equipped with a GPS) and transmits it to the base station;
  \item \relax Calculate a collection of k nearest neighbors for each sensor node in the network. It should be noted that the meanings of neighboring nodes are nodes that are closer to the node in terms of Euclidean distance;
  \item \relax Calculate pairwise distances between each node, and those nodes in its top-k nearest neighbor list for which the node is also on the neighbor’s nearest neighbor list, following equation (\ref{vequ8});
  \item \relax Merge all pairs of nodes with distances below a threshold.
  \end{enumerate}

\begin{figure}[!ht]
\centering
\includegraphics[width=8cm]{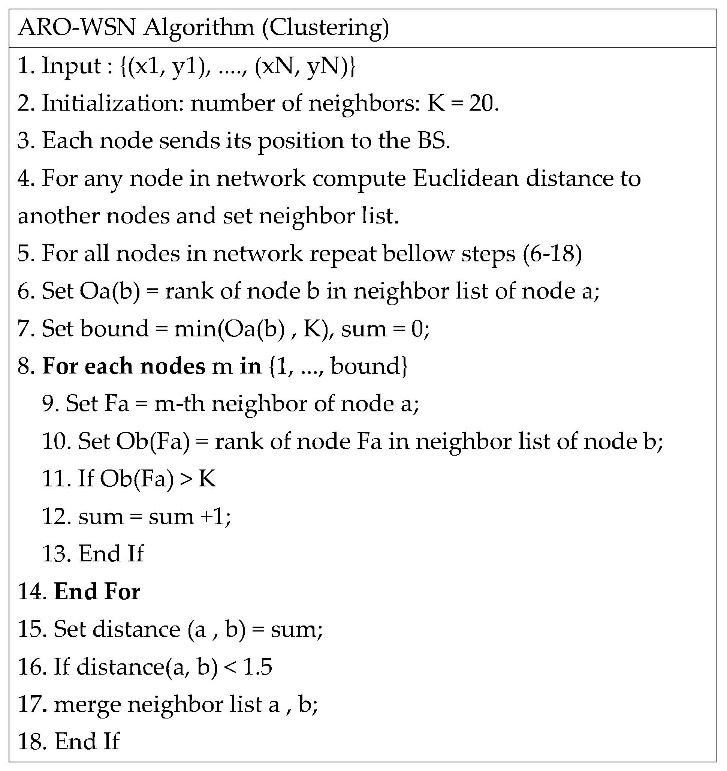}
\caption{The pseudo code of the ARO-WSN clustering step}\label{vfig2}
\end{figure}

The pseudo code of the clustering step of our proposed approach, which is referred to below as ARO-WSN, is depicted in Figure \ref{vfig2}. In our proposed approach, for the N set of sensor nodes (the value of N varies from 100 to 300 nodes in experiments), the k parameter is considered to be 20. This value has been selected experimentally after several experiments. The value of the threshold for equation (\ref{vequ8}) is also set to 1.5, according to the test in Section 5-4.

\subsubsection{Cluster head selection step}
\begin{figure}[!ht]
\centering
\includegraphics[width=8cm]{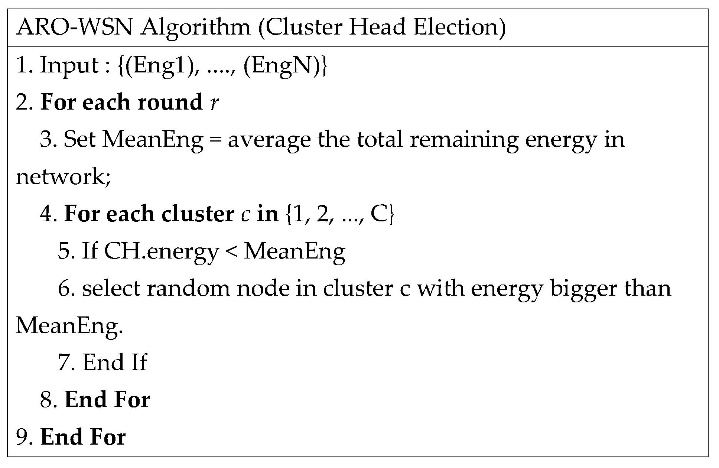}
\caption{The pseudo code of the ARO-WSN cluster head selection step}\label{vfig3}
\end{figure}

After the cluster is set, the next step is to select the cluster head. As previously emphasized, the clustering step is performed only once before executing repetitions, and is the only cluster head to be replaced in each round. In the proposed approach, the selection of cluster heads is random and dependent on the residual energy of the nodes. The selection of cluster heads is such that in each round, it is first tested whether or not the current cluster head energy is greater than the average total energy of the network. If the energy of the cluster head is greater than the average energy of all nodes, it can also be in the next round and it does not give its role to the other node. But if its energy is less than the average energy of the nodes, then randomly one of the other nodes is candidate, and after checking its energy level, if it has the necessary energy, is selected as the cluster head. This node should send an advertisement message with a node identifier in order to be new cluster head in the r round. As a result, each cluster head will be able to collect data from cluster nodes and transfer the collected data to the base station. The pseudo code of the cluster head selection step in the ARO-WSN algorithm is shown in Figure \ref{vfig3}.

\subsubsection{Data transmission}
\begin{figure}[!ht]
\centering
\includegraphics[width=8cm]{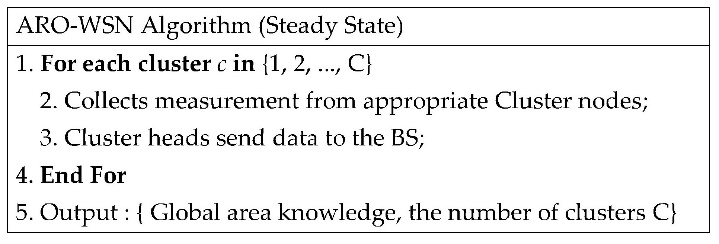}
\caption{The pseudo code of the ARO-WSN data transmission step}\label{vfig4}
\end{figure}

Once clusters and cluster heads are created, each normal node sends the sensed data to its designated cluster head. After that all data is received, each cluster head sends the collected data directly to the base station. The pseudo code of the data transmission step in ARO-WSN algorithm is shown in Figure \ref{vfig4}.

\section{SIMULATION RESULTS}
We have done extensive simulations based on the following setup to evaluate our proposed solution. We consider a dense wireless sensor networks with N nodes which are randomly distributed in a $100m\times100m$ field. The base station is located outside of the sensing area at a location with coordinates $(x_{sink}=0.5*x,y_{sink}=0.75*y+y)$. The number of nodes in all experiments is constant and equal to 500, except for scenario in which we test the effect of node density. Since nodes have limited energy, their energy levels is decreasing during the simulation period. When the energy storage of a node is evacuated, it is considered as a dead node and so, it cannot transmit or receive data \cite{9}. In all simulation scenarios, transmission or reception of data makes energy consumption. The parameters used for the energy model of all three algorithms in the simulations can be found in Table \ref{vtab1}.

\begin{table}[!ht]
\centering
\caption{The parameters of the radio model used in this simulation}\label{vtab1}
\begin{tabular}{|c|c|c|}
\hline
Parameter & Value \\ \hline
$E_{elec}=E_{Tx}=E_{Rx}$ & 50 nJ/bit\\
$E_{fs}$ & 10 pJ/bit/$m^2$\\
$E_{mp}$ & 0.0013 pJ/bit/$m^4$\\
$E_{DA}$ & 5 nJ/bit/msg\\
$d_o$ & 88 m\\
Message size & 4000 bits\\

\hline
\end{tabular}
\end{table}

In the following, our proposed ARO-WSN algorithm is compared with the following three algorithms:

\begin{itemize}
  \item \relax The K-means protocol is a distance-based clustering algorithm. Since our proposed approach is also a distance-based algorithm, K-means is a good choice for the purpose of comparison.
  \item \relax The LEACH protocol is one of the most famous hierarchical clustering protocols in WSNs. Since our proposed approach adopts a hierarchical approach, we will compare ARO-WSN with LEACH.
  \item \relax The LEACH-C protocol is a centralized, energy-aware extension of the LEACH protocol. Since the ARO-WSN algorithm is also a centralized and energy-aware approach, LEACH-C is a good candidate for the evaluations.
  
  \item \relax LEACH protocol with fuzzy descriptors.
\end{itemize}

Our criteria for the network lifetime is First Node Dead (FND); the moment when the battery storage of first sensor node in the network is completely discharged and Last Node Dead (LND); the moment when the battery of all sensor nodes in the network is completely depleted. Simulation results are the multi-run average of a random, constant size topology for all the algorithms.

\subsection{Impact of node density}
\begin{figure}[!ht]
\centering
\includegraphics[width=10cm]{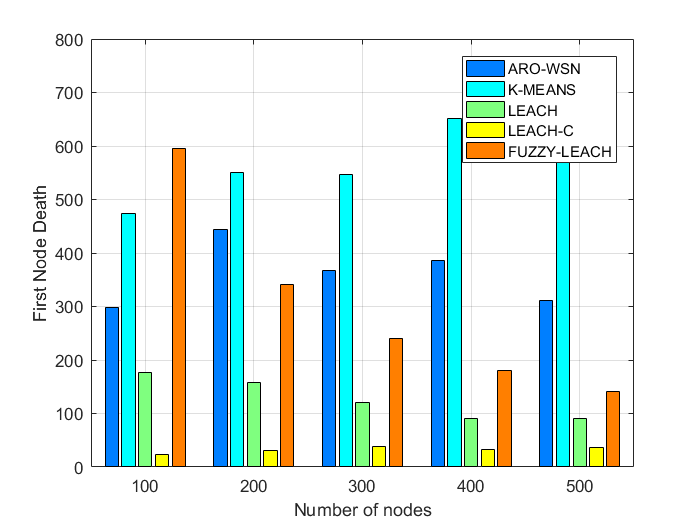}
\caption{Impact of node density on performance of compared algorithms with FND criterion}\label{vfig5}
\end{figure}

\begin{figure}[!ht]
\centering
\includegraphics[width=9.5cm]{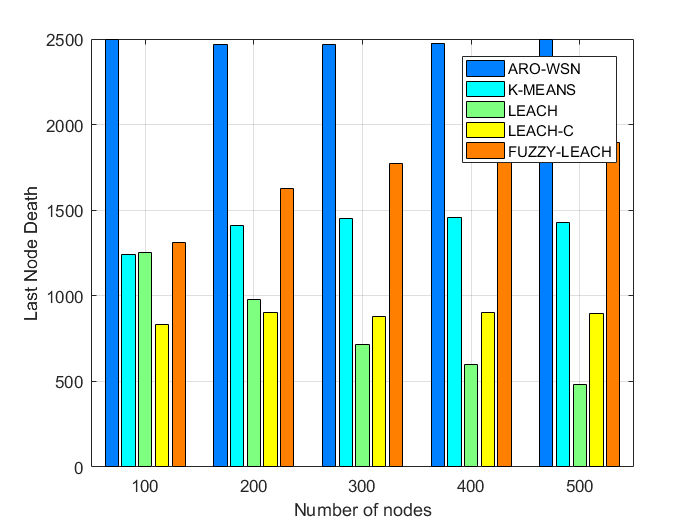}
\caption{Impact of node density on performance of compared algorithms with LND criterion}\label{vfig6}
\end{figure}

In this section, we evaluate the lifetime of the wireless sensor network for the different values of N based on FND and LND metrics. Figure \ref{vfig5} and Figure \ref{vfig6} show the effects of node density on the network lifetime, with FND and LND metrics in comparable clustering methods, respectively.

As shown in Figure \ref{vfig5}, for different values of N from 100 to 500, the ARO-WSN algorithm shows a performance improvement compared to the LEACH and LEACH-C algorithms. But the ARO-WSN algorithm has failed to improve its performance compared to the K-means algorithm. Despite the better performance of k-means in this case, the proposed algorithm is a better option for clustering sensor networks due to the flexibility in the number of clusters and less time overhead. From the value N = 200, the ARO-WSN algorithm has a greater FND value than FUZZY-LEACH. Unlike the ARO-WSN, it seems that the FUZZY-LEACH algorithm loses its efficiency at higher densities. This also applies to the LEACH and LEACH-C algorithms. We will further analyze the details of the chart.

In Figure \ref{vfig5}, the LEACH-C protocol has shown worse performance than LEACH. This is due to the fact that, in homogeneous environments, the formation of the clusters and the selection of the cluster heads in a decentralized way is more reasonable. In homogeneous network settings, all nodes have the same energy levels and if the cluster formation is done centralized, cluster heads must transfer all the new clustering information to the base station in each round with the same capabilities as the normal nodes. Although the ARO-WSN algorithm is also a centralized algorithm, the clustering phase in this algorithm is performed only once just before choosing the cluster heads. In ARO-WSN, the location of all nodes and members of each cluster is sent only once to the base station, and this information will not change during the period. Only once the cluster head is updated, and if the energy level of current cluster head is lower than the average total energy of the network, the advertisement message will be broadcast. Generally, the ARO-WSN and K-mean algorithms have better performance than LEACH and LEACH-C. Compared to LEACH, ARO-WSN and K-means algorithms, in addition to having the advantages of hierarchical clustering, perform based on distance measure. As shown in the energy consumption model in equation (\ref{vequ2}), energy consumption depends directly on the distance between the nodes and the cluster head and the distance among the cluster heads and the base station. Therefore, distance-based approaches consume less energy than LEACH, because one of the disadvantages of LEACH is the possibility to select inappropriate cluster heads in terms of location. If the LEACH-C algorithm is compared in a heterogeneous environment with other algorithms, it will perform better; but in the homogeneous environment, it has poor performance for the reasons explained. The ARO-WSN algorithm performs better at higher densities than FUZZY-LEACH and experiences the first node death in a longer time period than FUZZY-LEACH. Unlike the ARO-WSN, FUZZY-LEACH creates a new cluster in each round, which requires more energy. The disadvantages of this work are not shown at lower density, but in higher densities requiring more information to be transmitted, the advantage of forming clusters before the start of repetitions is clearly seen. The ARO-WSN with the first node death criterion has improved about \%60 compared to LEACH, about \%85 compared to LEACH-C and about \%22 compared to FUZZY-LEACH in all densities.

In Figure \ref{vfig6}, we investigate the impacts of node density on network lifetime with LND criterion in compared clustering methods.

Although in the FND criterion, the ARO-WSN algorithm failed to the K-means algorithm and in the low density to the FUZZY-LEACH algorithm, with the LND criterion, it was able to achieve good improvement over all four other algorithms. In more number of rounds, the advantage of performing clustering before selecting the cluster head and fixing the clusters more visible. The proposed approach with the last node death criterion has improved by \%42, \%67, \%64 and \%24, respectively, against the K-means, LEACH, LEACH-C and FUZZY-LEACH algorithms. As shown in Figure \ref{vfig6}, the ARO-WSN algorithm has almost constant LND value at different densities of the node, indicating that this approach has good stability and scalability at different densities of the node.

\subsection{Energy and lifetime}
\begin{figure}[!ht]
\centering
\includegraphics[width=10cm]{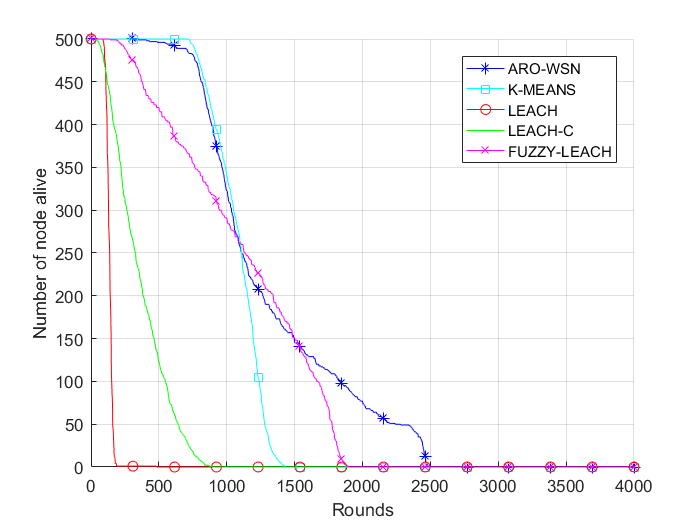}
\caption{Number of alive nodes over time of the compared protocols}\label{vfig7}
\end{figure}

In these experiments, we run compared algorithms under many networks with N = 500 nodes. The primary energy of all nodes is equal to 0.5J. The amount of data to send to the base station is unlimited. Figure \ref{vfig7} shows the number of active nodes in time for the ARO-WSN, K-means, LEACH, LEACH-C and FUZZY-LEACH algorithms.

As can be seen clearly in Figure \ref{vfig7}, the ARO-WSN algorithm with both FND and LND criterion has significantly improved in relation to the LEACH, LEACH-C and FUZZY-LEACH algorithms. In comparing two K-means and ARO-WSN algorithms with the FND criterion, the ARO-WSN algorithm has a relatively large distance to the K-means curve.

\begin{figure}[!ht]
\centering
\includegraphics[width=10cm]{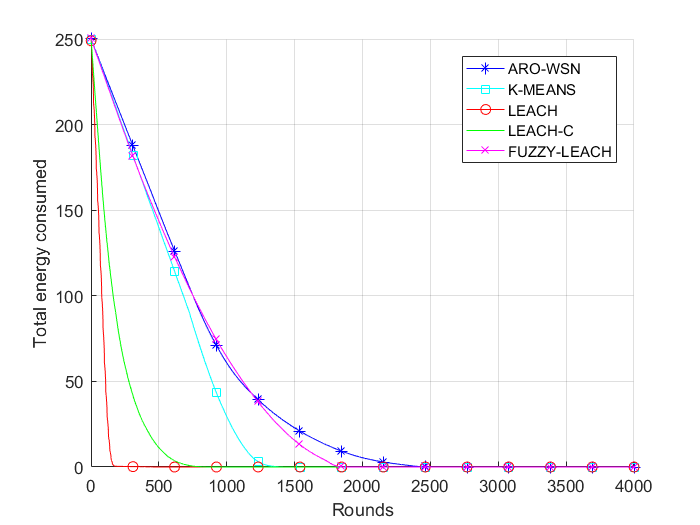}
\caption{The evolution of the remaining energy in the network}\label{vfig8}
\end{figure}

In Figure \ref{vfig8}, we observe the total residual energy of the network in each round of transmission using compared approaches. This chart clearly and simply depicts the increasing network lifetime. In this diagram, the LEACH algorithm is the first approach in which the total energy of the network ends, and then the LEACH-C algorithm has the second rank in the discharge of network energy. The K-means algorithm, although with a great distance from the LEACH-C algorithm approaches the completion of the entire network energy, about 1,000 rounds faster than the ARO-WSN algorithm experiences the death of the last node and the total energy depletion of the network. The death of the last node with the FUZZY-LEACH approach is also 600 rounds faster than the proposed approach. Figure \ref{vfig8} illustrates the fact that the ARO-WSN algorithm has been successful in increasing network lifetime.

The ARO-WSN algorithm has consumed only \%38 of the total network energy in round 500. While K-means, LEACH, LEACH-C and FUZZY-LEACH algorithms have consumed \%44, \%100, \%95.6 and \%42 of total energy, respectively. At round 1500, the K-means, LEACH, and LEACH-C algorithms have consumed all of the network's energy, but the ARO-WSN algorithm has stored about \%10 of the total energy, and can still continue to collect information from the environment around the network. In the 1900s, the FUZZY-LEACH algorithm lost all of the network's energy, and the ARO-WSN algorithm remained at \%3 of the energy in the network.

\begin{figure}[!ht]
\centering
\includegraphics[width=10cm]{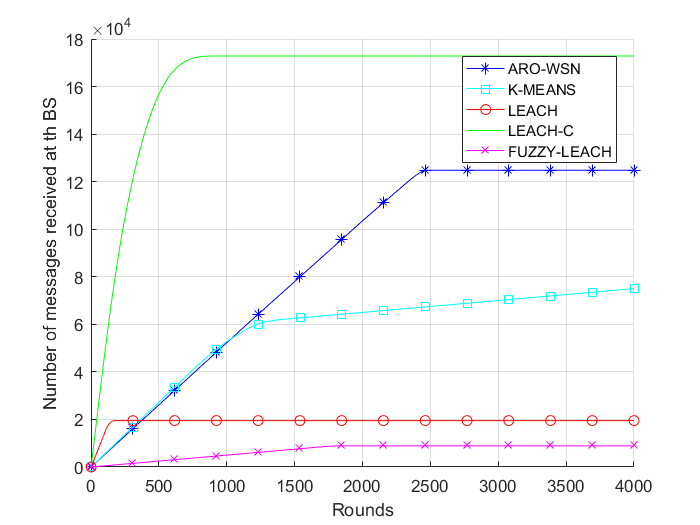}
\caption{Number of messages received at the base station over time}\label{vfig9}
\end{figure}

Figure \ref{vfig9} shows the number of messages received by the base station for the compared approaches.

In Figure \ref{vfig9}, it has been shown that in each algorithm, the incremental trend of transmission of messages in the corresponding curve is fixed at a specific time moment. This means that at that moment, the last node in the network failed and the network lifetime has been over. In this diagram, the LEACH and LEACH-C algorithms initially transmitted more messages to the base station. These messages are a collection of advertisement messages and data collected by cluster heads. In the implementation of these algorithms, channel failure issues are not considered; therefore, all nodes can transfer data collected from the environment to the cluster head. It can be concluded that the difference in the number of messages transmitted to the base station is related to the advertisement messages broadcasted by the LEACH and LEACH-C algorithms. According to equation (\ref{vequ2}), the amount of energy consumed depends on the size of the messages; therefore, the greater the number of messages sent and received in the sensors, the greater the power consumption in the network. The faster transmission of the message in the LEACH and LEACH-C algorithms has led to faster energy consumption in both of these approaches, resulting network lifetime faster has finished. Here, the lower number of advertisement messages has caused superiority of the ARO-WSN algorithm, which indeed is the result of the lack of repeating the clustering in each round. The K-means curve is close to the ARO-WSN message transmission curve. This shows that the algorithm acts more cautiously in transmission of messages than two LEACH and LEACH-C algorithms; but as shown in Figure \ref{vfig9}, the moment of the death of the last node in this algorithm is far from the ARO-WSN algorithm; therefore, there is no more opportunity to transfer message packs to the base station.

\begin{figure}[!ht]
\centering
\includegraphics[width=10cm]{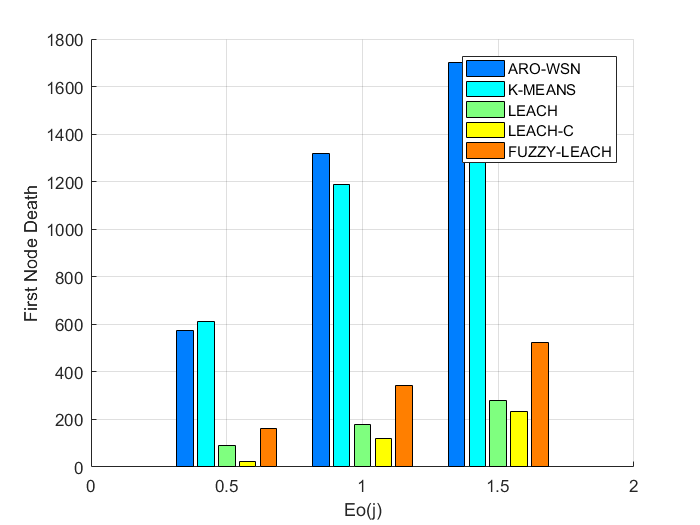}
\caption{Impact of the initial energy amount on the performance of compared algorithms with FDN criterion}\label{vfig10}
\end{figure}

\begin{figure}[!ht]
\centering
\includegraphics[width=10cm]{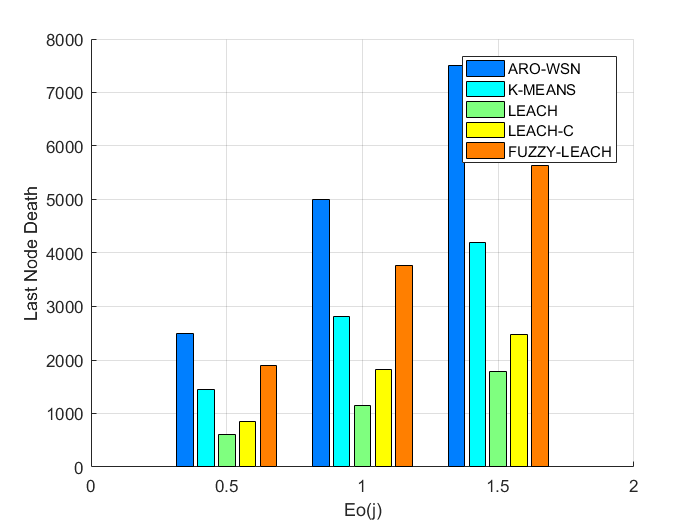}
\caption{Impact of the initial energy amount on the performance of compared algorithms with LDN criterion}\label{vfig11}
\end{figure}

Figure \ref{vfig10} and Figure \ref{vfig11} show the performance of the compared algorithms with different primary energies with FND criterion and LND criterion, respectively.

According to Figure \ref{vfig10}, we see that the performance of all the compared algorithms is improved by increasing the initial energy. The ARO-WSN algorithm works better than other algorithms with increasing initial energy. Therefore, it can be concluded that distance partnership in the formation of a cluster can be much more effective than the initial energy level. With increasing primary energy, the difference in FND value between K-means and ARO-WSN distance-based algorithms increases with the two LEACH and LEACH-based algorithms. With the initial increase in primary energy, the proposed approach, which did not have the FND score superior to the K-means algorithm, experiences the first death later than the K-means algorithm. Figure \ref{vfig11} gives similar results to Figure \ref{vfig10}, with the difference that here the difference the LND value between the two ARO-WSN and K-means algorithms is also significant. 

\subsection{Impact of message size}
In all experiments that have been done so far, the size of the messages sent from the nodes to the cluster head and from the cluster head to the base station is constant and equal to 4000 bits. According to Equation (\ref{vequ2}), the size of the messages is effective in the energy consumption model for sending and receiving messages; therefore, the larger the size of messages sent and received in the sensor nodes, the more energy is consumed and the network lifetime decreases. In this experiment, we want to show the effect of the size of received and sent messages from sensor nodes in the performance of the compared approaches.

\begin{figure}[!ht]
\centering
\includegraphics[width=10cm]{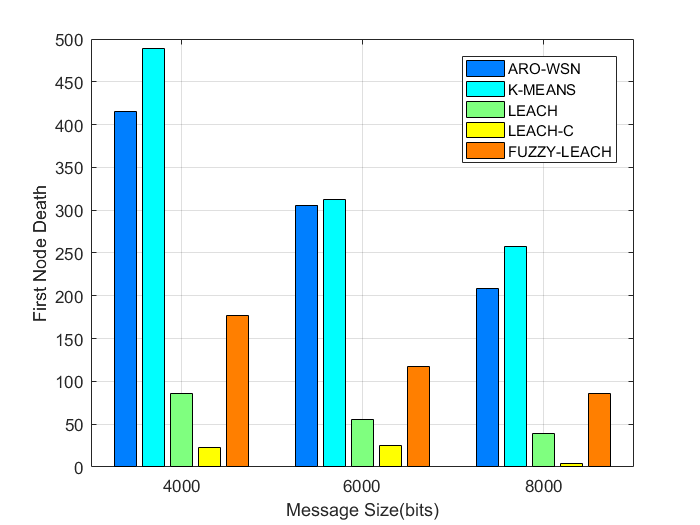}
\caption{Impact of message size on the performance of compared algorithms with FDN criterion}\label{vfig12}
\end{figure}

\begin{figure}[!ht]
\centering
\includegraphics[width=10cm]{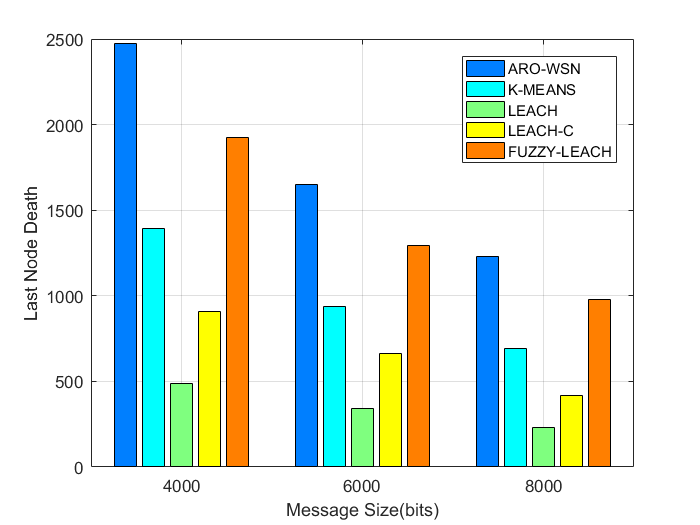}
\caption{Impact of message size on the performance of compared algorithms with LND criterion}\label{vfig13}
\end{figure}

Regarding Figure \ref{vfig12} and Figure \ref{vfig13}, with increasing message size, the first and last node death occurs in all approaches earlier, but the ARO-WSN algorithm still maintains its improvement over other algorithms. Figure \ref{vfig12} shows that in the message size of 4000 bits, the ARO-WSN algorithm has a lower FND value than K-means. This issue is also present in all the graphs that have been presented so far and the proposed algorithm has not been able to improve with the FND criterion toward to the K-means algorithm. As previously mentioned, the clustering step in the ARO-WSN algorithm is performed before selecting the cluster head. Also, cluster head update is only performed if the cluster head energy is less than the average total energy of the network; but in the other three approaches, clustering is performed in each round. The messages exchanged in each round to create new clustering and cluster head updates in these three approaches are greater than the messages sent to the base station in the ARO-WSN algorithm for updating cluster heads (see Fig. \ref{vfig9}); therefore, according to equation (\ref{vequ2}), by increasing the size and number of messages sent and received, more energy is consumed and the lifetime of the network decreases.

\subsection{Effect of parameter C value}
\begin{figure}[!ht]
\centering
\includegraphics[width=10cm]{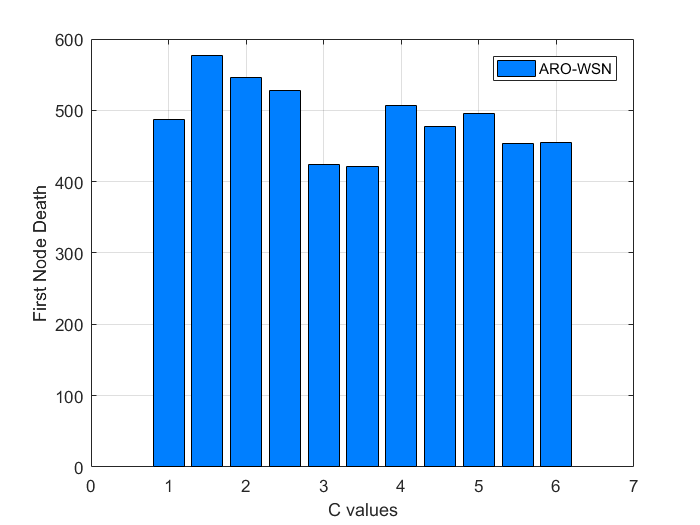}
\caption{The Impact of parameter C on the performance of the proposed approach with the FND criterion}\label{vfig14}
\end{figure}

As indicated in Section 4.2, selecting the threshold C for determining the number of clusters in a dataset is one of the most difficult issues in data clustering. In practice, it cannot be assumed that the actual number of clusters is predetermined, so this algorithm is evaluated in several effective values of C and the best result is reported. In Section 4.3.1 this value is considered for the proposed approach to 1.5. Figure \ref{vfig14} shows the effect of the value of the parameter C on the proposed approach with the death of the first node criterion. As shown in Figure \ref{vfig14}, the ARO-WSN algorithm has a greater FND value in C = 1.5 and for all experiments, 1.5 is selected for this parameter in the proposed approach.

\section{CONCLUSION}
The purpose of this paper is to increase the lifetime of wireless sensor networks through clustering. Given the hierarchical and distance-based clustering features, the ARO algorithm, which is a hierarchical clustering based on distance and has been used in the image processing to clustering millions of images, was selected in order to clustering network sensors. In addition to using the ARO algorithm at the clustering step, the idea of performing the clustering step before choosing the cluster head was also used in the proposed ARO-WSN approach. This idea will reduce the network power consumption by eliminating overhead advertisements per round and thus increase network lifetime. The simulation results show that in the ARO-WSN, the death of the last node compared to the LEACH, LEACH-C, K-means and FUZZY-LEACH algorithms, and the death of the first node compared to the LEACH and LEACH-C algorithms, and in the higher densities than the FUZZY-LEACH algorithms occurs later. The ARO-WSN with the first node death criterion has improved about \%60 compared to LEACH, about \%85 compared to LEACH-C and about \%22 compared to FUZZY-LEACH in all densities. In addition, with the last node death criterion was improved by \%42, \%67, \%64 and \%24, respectively, with the K-means, LEACH, LEACH-C and FUZZY-LEACH algorithms.
In terms of energy consumption, when the K-means, LEACH and LEACH-C algorithms consume all the network energy, the proposed approach saves \%10 of the total network energy and can still continue to collect information from the surrounding environment. Therefore, the proposed ARO-WSN approach has succeeded in reducing energy consumption through clustering and, consequently, increasing the lifespan of wireless sensor networks.

\nocite{*}% Show all bib entries - both cited and uncited; comment this line to view only cited bib entries;
\bibliography{wileyNJD-AMA}%

\end{document}